\documentclass[twocolumn]{aastex63}

\received{2024 August 5}
\revised{2024 November 22}
\accepted{2024 November 25}

\usepackage{enumitem} 
\usepackage{lipsum}

\usepackage{lineno}
\shortauthors{Blackwell et al.}
\graphicspath{{./}{figures/}}

\begin{document}

\defcitealias{Loew2013}{L13}

\title{Early Enrichment Population Theory at High Redshift}

\author[0000-0002-8195-0563]{Anne E. Blackwell}
\affiliation{University of Michigan, Ann Arbor, MI 48104, USA}

\author[0000-0001-6276-9526]{Joel N. Bregman}
\affiliation{University of Michigan, Ann Arbor, MI 48104, USA}











\begin{abstract}

An Early Enrichment Population (EEP) has been theorized to produce the observed intracluster medium (ICM) metallicity of galaxy clusters. 
This population likely existed at high redshifts (z$\sim$10), relics of which we posit exist today as dwarf galaxies. 
Previous work argues that the initial mass function (IMF) of the EEP must be flatter than those found at lower redshifts, but with considerable uncertainties. 
In this work, we present a more quantitative model for the EEP and demonstrate how observational constraints can be applied to the IMF using supernova Type Ia (SN Ia) rates, delay time distribution (DTD), and the luminosity function (LF) of galaxy clusters. 
We determine best-fit values for the slope and mass break of the IMF by comparing IMFs from literature with observed DTDs, and the low-luminosity component ($M(R)> -12$) of the Coma LF. We derive two best-fit IMFs, flatter than standard IMFs: (1) $\alpha_{lo} = -0.13 \pm 0.24$ for $0.07 < M/M_\odot < 1.75$ and $\alpha_{hi} = 0.53 \pm 0.01$ for $1.75 < M/M_\odot < 150$, and (2) $\alpha_{lo} = 1.06 \pm 0.11$ for $0.07 < M/M_\odot < 6$ and $\alpha_{hi} = 0.53 \pm 0.01$ for $6 < M/M_\odot < 150$. 
We compare these with \textit{sl-5} from \cite{Loew2013}, with $\alpha=0.5$ for $0.07 < M/M_\odot < 8$, and $\alpha=0.3$ for $8 < M/M_\odot < 150$. 
This EEP model, along with stars formed at later times, can produce the observed ICM metallicity, is consistent with observations, and predicts a significant rise in the SN Ia rate at increasing redshift.

\end{abstract}



\section{Introduction} \label{sec:intro}
Galaxy clusters are the largest gravitationally bound structures in the Universe. 
Their deep gravitational potential wells allow us to approximate the largest clusters ($10^{13.5} - 10^{15} M_\odot$) as closed-box systems - all elements produced within the cluster remain there.
This makes the hot, X-ray emitting halo the perfect place to study the elemental history of galaxy groups and clusters.
The X-ray halo lies within the virial radius of the galaxy cluster, and is seen in the energy range $\sim$0.5-12.0 keV - the intracluster medium (ICM).

Many have studied the gas phase metallicity of clusters \citep[e.g.,][]{Baldi2012, Mantz2017, Ghiz2020, Lovisari2020} and find a universal metallicity $\sim$0.35 $Z_\odot$ outside 0.1$R_{500}$ using \cite{Asplund2009} abundances \citep[][]{deGrandi2004, Mantz2017, Urban2017}.
The ICM metallicity ($Z_{ICM}$) seems to be independent of galaxy cluster mass, or stellar fraction ($M_*/M_{gas}$), even over a large range of $M_*/M_{gas}$ (0.035-0.43) \citep{Breg2010, Blackwell2022}. 
This is especially confounding when the measured halo metallicity is compared to expectations.
\cite{Loew2013} developed a trend to determine the halo metallicity from the visible stellar populations, $Z_*=1.55(M_*/M_{gas})$.
For a cluster with $M_*/M_{gas}=0.05$, the expected halo metallicity is 0.08 $Z_\odot$, a factor of $>$4 below $Z_{ICM} \sim 0.35 Z_\odot$.
That is, the visible galaxies and stars in a galaxy cluster cannot have produced the measured ICM metallicity.
This discrepancy is the missing metal conundrum, and has been discovered and discussed previously \cite[e.g.][]{Elbaz1995, Mush1996, Port2004, Renzini2014, Werner2020}.

Various potential explanations for the missing metal conundrum have been explored, with fault found in each.
One solution proposes the presence of stars in the ICM, which generated sufficient quantities of metals to account for the observed metallicity in high-mass clusters \citep{Renzini2014}.
\cite{Siv2009} calculated that intracluster stars with a standard initial mass function (IMF) contribute approximately 25\% of the total ICM metallicity within $R_{500}$, but only 20\% of the total intracluster light (ICL) \citep{Zibetti2005, Krick2007, Giallongo2014, Furnell2021}.
If these intracluster stars were to supply the missing metals, we would expect to observe a population of intracluster stars at least four times larger than what is detected.
A population four times larger, sufficient to explain the absence of metals, would generate at least 80\% of the total cluster light, exceeding observations.

Another theory examines the relationship between the IMF of star-forming clouds at $z\sim 2$ and cluster final mass \citep{Renzini2014}.
High-mass clusters require an additional source of metals that cannot be accounted for by visible stellar populations. 
These metals might originate from star-forming clouds at high redshifts prior to the formation of galaxy clusters.
To compensate for the missing metals, the IMF of star-forming clouds at high redshifts would need to produce more high-mass stars for higher final mass systems, and fewer high-mass stars for lower final mass systems \citep{Renzini2014, Blackwell2022}.
This scenario assumes the star-forming clouds ``know" the final cluster mass, which is improbable.

We are left with one theory to consider, a population of stars that existed at high redshifts and supplied high-mass clusters with the missing metals - the early enrichment population (EEP) \citep{Elbaz1995, Loew2001}.
This theory has emerged from investigations of the evolution of halo metallicity across redshift \citep{Mantz2017, Flores2021}, and supported by studies of metallicity in relation to stellar fraction \citep{Breg2010, Blackwell2022}.
According to this theory, the EEP would have existed during high redshifts, primarily consisting of early Population II stars.
These stars would have played a significant role during the reionization period in protoclusters, $6 < z < 10$, and would have favored a bottom-light IMF \citep{Fan2006, Loew2013, Planck2016reionization}.
The EEP would have been present in nascent galaxies, not prominent as $L^*$ galaxies, and would not be observable today \citep{Werner2020}.

Currently, very little is known about this theory. 
Few authors have provided calculations or predictions for the contribution of the EEP (\citealt{Loew2013}, hereafter \citetalias{Loew2013}).
\cite{Blackwell2022} recently made calculations of observable and testable quantities for the EEP, such as the expected Type Ia supernovae rate.
In this paper, we examine further aspects about the current state of the EEP theory (Section \ref{sec:eep}), and provide calculations of predictable and measurable quantities to test this theory, such as supernova Type Ia (SN Ia) observations (Section \ref{sec:SNIa}) and Luminosity Functions (LF, Section \ref{sec:lf}).
These calculations are built off of predictions of $Z_{EEP}$, Section \ref{sec:survey}.
The assumptions required to make these calculations and are discussed in Section \ref{sec:assumptions} and Section \ref{sec:imf}.

\section{The Early Enrichment Population}
\label{sec:eep}
 
The idea of early chemical enrichment has been discussed for the past three decades \cite[e.g.,][]{Renzini1993, Elbaz1995, Mush1996, Loew2001, Matteucci2005, Million2011, Mantz2017, Blackwell2022, Sarkar2022}, since the first X-ray spectra of galaxy clusters were studied with an exceptionally prominent Iron (Fe) emission line \citep[e.g.,][]{Gursky1971, Mitchell1976, Mush1978}.
There are small variations, but the general idea between all theories is consistent: the excess abundance of metals in the ICM of high-mass clusters, and a seemingly Universal metallicity among galaxy groups and clusters ($\sim$0.35$Z_\odot$), are not well explained by standard galaxy cluster enrichment theories. 
Galaxies within a rich cluster do not have a large enough stellar population to have enriched the ICM to the measured abundance using a fixed IMF. 
A universal metallicity suggests that an earlier, and external source of metals must exist.

The early enrichment population (EEP) is a theorized population of stars that created most of the metals in high-mass galaxy clusters and smaller contributions in low-mass galaxy clusters that are measured in the ICM today.
This population would be primarily early Pop II and late Pop III stars.
The exact IMF for Pop III stars remains uncertain; however, it is anticipated to be dominated by medium and high-mass stars ($10-10^3 M_\odot$) \citep[][and references therein]{Abel2002, Bromm2002, Bromm2013, Chantavat2023, Klessen2023}.
Supernovae resulting from the core collapse of high-mass stars (SNcc) primarily yield alpha elements (i.e., carbon, oxygen, nitrogen, magnesium and silicon), and a subset of metal group elements. 
Pair instability supernovae can occur in stars with masses $\sim$100-260 $M_\odot$. 
The amount of Fe they produce is negligible, except at the high mass end ($\gtrsim$150 $M_\odot$), even if they are relatively common \citep{TNY1986, TNH1996, Ferreras2002, Heger2005, Woosley2015}. 
The majority of Fe observed in galaxy clusters is generated by SN Ia with a typical yield of 0.743 $M_\odot$ per SN Ia, and 0.0824 $M_\odot$ per SNcc \citep{Kobayashi2006, Loew2013}.

Once Pop III stars pollute their surroundings to a metallicity level exceeding $10^{-6}-10^{-3.5}$ $Z_\odot$, the cooling process is primarily governed by fine structure lines \citep{Bromm2003, Fang2004, Bromm2013, Pall2014, Yang2015, LiuBromm2020, Corazza2022}.  
The IMF is then believed to adopt forms similar to those observed during the standard low-redshift regime, marking the transition to the Pop II phase.
The Pop II phase overlaps with Pop III, but swiftly surpasses it in terms of significance for metal or photoionization production at redshifts z$\sim$11-13 \citep{Yang2015, Corazza2022}.

Before reionization, the gas remains in a neutral state.
This enables easier collapse into stars, particularly in dwarf galaxies, or dwarf elliptical (dE) galaxies in clusters, which are a likely host for the EEP.
At $z\sim 10$, the $T_{CMB}\sim 30$ K, which raises the Jeans mass and potentially modifies the low-mass segment of the IMF.
During the reionization period ranging $6<z<9$, it is expected that Pop II stars dominate \citep{Fan2006, Planck2016reionization, Jaacks2019}.
For these reasons, we believe the EEP primarily occurs before reionization reaches completion, and is dominated by a bottom-light and/or top-heavy IMF.
Few works have investigated the exact shape of the EEP IMF, narrowing possibilities simply on the metal abundances required \citep{Loew2013, Corazza2022}.

Initial studies of the EEP contribution to galaxy groups and clusters suggests that the EEP may have begun at different epochs (z=12 vs z=7) \citep{Corazza2022}.
Systems with the same value of $M_*/M_{gas}$, similar halo masses, and implied EEP metallicity contribution ($Z_{EEP}$) have varying $Z_{ICM}$ around the mean (discussed further in Section \ref{sec:survey}).
This implies differences in the efficiency of the EEP in different galaxy clusters, which may be a result of some systems having longer periods of star formation with a flatter IMF.

Two further constraints on the EEP IMF are the observed SN Ia rate and the remaining light from the EEP.
First, more recent star formation would result in a higher occurrence rate of SN Ia at more recent times.
Second, stars with mass $<$1$M_\odot$ would still exist today if star formation occurred at $z\sim10$.
If the EEP formed pre-reionization, we would expect to see these stars in the dE galaxies in groups and clusters.
Thus, the number of SN Ia and total remaining luminosity of dE galaxies (the theorized hosts of the EEP) within a cluster constrain the high and low-mass end of the EEP IMF respectively, and the time of formation.

\section{$Z_{ICM}$ versus $M_*/M_{gas}$ Survey}
\label{sec:survey}

The main objective of this survey was to perform a comprehensive and uniform analysis of a large data set over a range of 0.035-0.43 $M_*/M_{gas}$ (Table \ref{tab:survey}).
We obtained values of $M_{*,500}$ and $M_{gas,500}$ for 25 systems from a single source to gain deeper insights into the relationship between $Z_{ICM}$ and $M_*/M_{gas}$ \citep{Lagana2013}.
Each system was identified to have sufficient counts ($>$40,000 within $R_{500}$) of archival \textit{XMM Newton} data to derive the average ICM metallicity ($Z_{ICM}$) with an error margin of approximately 10\%.
All systems have an angular size less than the \textit{XMM Newton} EPIC camera, allowing for an on-chip background to be used.
Additional systems were reported in \cite{Lagana2013} with suitable $M_{*,500}$, $M_{gas,500}$, and sufficient archival data. 
However, these systems were not relaxed and therefore not suitable for this survey (i.e., A115 \citealt{Gutierrez2005, Hallman2018}).

\begin{table*}[]
    \hspace{-1.5cm}
    \centering
    \begin{tabular}{c | c | c | c | c}
    \hline
    Cluster & RA & DEC & $M_*/M_{gas}$ & $Z_{ICM}$ $Z_\odot$ \\
         \hline
         A2069 & 15:24:39.8 & +29:53:26.3 & 0.035$\pm$0.009 & 0.44$\pm$0.04 \\
         A697 & 08:42:57.6 & +36:21:55.8 & 0.044$\pm$0.013 & 0.34$\pm$0.04  \\
         A665 & 08:30:57.36 & +65:50:33.36 & 0.046$\pm$0.014 & 0.43$\pm$0.09\\
         MS1455.0+2232 & 14:57:15.12 & +22:20:35.52 & 0.066$\pm$0.020 & 0.49$\pm$0.03\\
         A1763 & 13:35:18.24 & +40:59:59.28 & 0.069$\pm$0.021 & 0.40$\pm$0.05\\
         A1914 & 14:26:99.96 & +37:39:33.96 & 0.069$\pm$0.021 & 0.40$\pm$0.040 \\
         A2409 & 22:00:52.8 & +20:58:27.84 & 0.073$\pm$0.022 & 0.47$\pm$0.03 \\
         RXJ1720+2638 & 17:20:16.8 & +26:38:06.6 & 0.077$\pm$0.024 & 0.56$\pm$0.03 \\
         A773 & 09:17:53.04 & +51:43:39.72 & 0.083$\pm$0.025 & 0.38$\pm$0.05 \\
         A1413 & 11:55:18 & +23:24:17.28 & 0.088$\pm$0.026 & 0.44$\pm$0.01 \\
         A781 & 09:20:26.16 & +30:30:02.52 & 0.091$\pm$0.028 & 0.31$\pm$0.05\\
         A2261 & 17:22:27.12 & +32:07:56.28 & 0.095$\pm$0.029 & 0.44$\pm$0.04 \\
         A1689 & 13:11:29.52 & -01:20:29.68 & 0.098$\pm$0.029 & 0.38$\pm$0.02\\
         A267 & 01:52:42.14 & +01:11:41.29 & 0.106$\pm$0.032 & 0.45$\pm$0.06 \\
         AWM4 & 16:04:57 & +23:55:14 & 0.111$\pm$0.026 & 0.49$\pm$0.03 \\
         A383 & 02:48:03.43 & -03:31:45.87 & 0.123$\pm$0.038 & 0.49$\pm$0.04 \\
         A1991 & 14:54:30.2 & +18:37:51 & 0.163$\pm$0.045 & 0.50$\pm$0.04 \\
         NGC4325 & 12:23:06.7 & +10:37:16 & 0.193$\pm$0.054 & 0.36$\pm$0.02 \\
         MS0906.5+1110 & 09:09:12.72 & +10:58:32.88 & 0.209$\pm$0.071 & 0.48$\pm$0.07 \\
         A2034 & 15:10:12.48 & +33:30:28.08 & 0.234$\pm$0.069 & 0.37$\pm$0.05 \\
         A2259 & 17:20:10.08 & +27:39:03.28 & 0.273$\pm$0.084 & 0.40$\pm$0.03 \\
         NGC4104 & 12:06:39 & +28:10:27 & 0.337$\pm$0.089 & 0.45$\pm$0.06 \\
         RXCJ2315.7-0222 & 23:15:44.1 & -02:22:59 & 0.376$\pm$0.011 & 0.41$\pm$0.02 \\
         A586 & 07:32:20.16 & +31:37:55.92 & 0.430$\pm$0.134 & 0.40$\pm$0.03 \\
         NGC1132 & 02:52:51.9 & -01:16:29 & 0.484$\pm$0.098 & 0.31$\pm$0.01 \\
         
    \end{tabular}
    \label{tab:survey}
    \caption{Summary of our galaxy group and cluster survey. Values of $M_*$, and $M_{gas}$ within $R_{500}$ were obtained from \cite{Lagana2013}. $Z_{ICM}$ values were measured spectroscopically from X-ray data.}
\end{table*}

We followed the data reduction and analyses methods outlined in \cite{Blackwell2022}.
We used SAS (version 19.1.0)\footnote{https://www.cosmos.esa.int/web/xmm-newton/sas} for all data reduction, and the most recent calibration files, CALDB (version 4.9.3)\footnote{http://cxc.harvard.edu/caldb/}.
We analyze the ICM metallicity in five radial bins originating from the X-ray center.
The radial bins were defined as: $<0.15$ $R_{500}$, $0.15-0.3$ $R_{500}$, $0.3-0.55$ $R_{500}$, $0.55-0.75$ $R_{500}$ and $0.75-1$ $R_{500}$.
During spectral extraction and fitting, we took the point spread function into account by generating \textit{crossarfs}.
These redistribute the X-ray photons into the respective annuli where they physically originated, but may not be detected.
Additionally, we extracted two spectra from the PN chip to improve counting statistics, using PATTERN=0 and PATTERN=4 for a low- and high-energy spectrum, respectively.
We excluded the central annulus when calculating $Z_{ICM}$ and took a weighted average of the fit-determined metallicites from all other annuli. 
Presentation of results of each cluster will be given in an upcoming paper.


\begin{figure}
    \centering
    \includegraphics[width=1\linewidth]{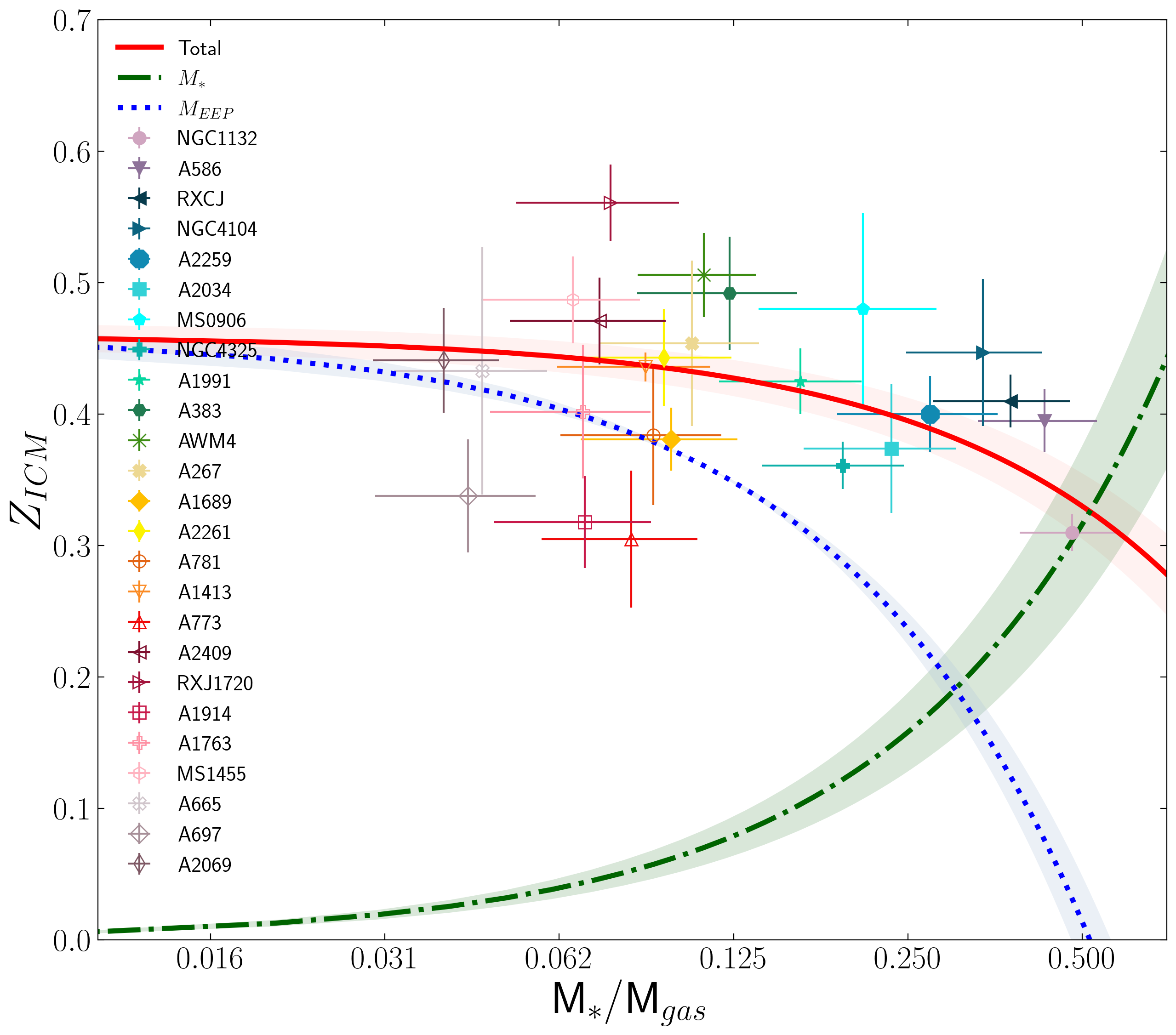}
    \caption{ICM metallicity measurements for 25 galaxy groups and clusters outside of the core of the cluster ($>$$0.15 R_{500}$) versus the stellar fraction, $M_*/M_{gas}$. The red solid line shows the total fit of $Z_{ICM}=(-0.26 \pm 0.03) \times (M_*/M_{gas}) + (0.46 \pm 0.01)$, done in linear-space. The x-axis is presented in log-scale. The green dash-dot line is the contribution from the stellar population, modeled as $Z_*=(0.63\pm0.12) (M_*/M_{gas})$. This fit is valid over $0.035 < M_*/M_{gas} < 0.43$. The trend for $Z_*$ is subtracted from the total fit to find $Z_{EEP}=(-0.89 \pm 0.12)(M_*/M_{gas}) + (0.46 \pm 0.01)$, shown in the dotted blue line. Shaded regions around each line are 1$\sigma$}.
    \label{fig:survey}
\end{figure}

The results of this survey are shown in Figure \ref{fig:survey} as a plot of $Z_{ICM}$ versus $M_*/M_{gas}$, and Table \ref{tab:survey}.
The overall fit to the trend was done using an MCMC to perform linear regression, minimizing the sum of the squares of the residuals.
The result is presented in Equation \ref{eq:tot_fit}, and is shown as the red line in Figure \ref{fig:survey}. 
\begin{equation}
Z_{ICM}=(-0.26\pm0.03)(M_*/M_{gas})+(0.46\pm0.01) 
\label{eq:tot_fit}
\end{equation}

This fit indicates a minor dependence of $Z_{ICM}$ on $M_*/M_{gas}$; however, we note the slope is driven by the data points at high $M_*/M_{gas}$.
There are only 7 data points with $M_*/M_{gas}>0.20$. 
We performed Jackknife resampling to investigate the significance of these data points.
Excluding the first data point alters the fit to $Z_{ICM}=(-0.15\pm0.05)(M_*/M_{gas})+(0.44\pm0.01)$.
The slope remains constant within $1\sigma$ removing subsequent highest $M_*/M_{gas}$ systems.
Galaxy cluster NGC1132 has the highest stellar fraction ($0.48\pm0.09$) and $Z_{ICM}=0.31\pm0.01 Z_\odot$, making it a powerful weight for the slope of $Z_{ICM}$ vs $M_*/M_{gas}$.
However, $Z_{ICM}$ of NGC1132 is still within the dispersion of $Z_{ICM}$ around the mean.
For that reason, we do not exclude it from the sample and adopt Equation \ref{eq:tot_fit} as the trend of $Z_{ICM}$.

We also note the dispersion of $Z_{ICM}$, the largest of which occurs at low $M_*/M_{gas}$.
We can make a rough estimate of the contribution of the intrinsic variation to the total scatter.
The total scatter may be written as $\sigma = \sqrt{\sigma_{met}^2 + \sigma_{int}^2}$, where $\sigma_{met}$ is the average error on $Z_{ICM}$, and $\sigma_{int}$ is the inferred intrinsic scatter.
The average error of $Z_{ICM}$ is 0.039 $Z_\odot$, and $\sigma$=0.052 $Z_\odot$ over all stellar fractions.
Solving for $\sigma_{int}$, we find 0.034 $Z_\odot$.
Approximately half of the scatter around the expected $Z_{ICM}$ is intrinsic.

The value of $Z_{ICM}$ is the sum of $Z_*$ and $Z_{EEP}$. 
These variables are crucial to this work, thus we clarify them here:

\indent $Z_{ICM}$ - the gas phase metallicity expelled from $Z_*$ and $Z_{EEP}$, measured spectroscopically from the X-ray data outside in regions $> 0.15 R_{500}$.

\indent $Z_{EEP}$ -  EEP metal abundance measured in the ICM. The optical remnants of the EEP do not contribute greatly to the present-day luminosity of clusters, nor are theorized to contribute primarily to the dwarf Elliptical galaxy cluster members.

\indent $Z_*$ - metal abundance measured in ICM produced by the visible stellar populations. These remnants dominate the luminosity in cluster members today.

Both $Z_*$ and $Z_{EEP}$ may be written as functions of their relative population stellar fractions, multiplied by a constant, as seen in Equation \ref{eq:metallicity}.

\begin{equation}
    Z_{ICM} = Z_* + Z_{EEP} = a_1 (M_*/M_{gas}) + a_2 (M_{EEP}/M_{gas})
\label{eq:metallicity}
\end{equation}

\cite{Loew2013} developed a relationship for $Z_*$ as a function of $M_*/M_{gas}$, specifically for Fe return, $Z_*=1.55(M_*/M_{gas})$.
This was determined by considering the metals locked up in stars, mixing of metals with the ICM, and efficiency of star formation. 
However, this derived trend of $Z_*$ overpredicts the amount of metals produced in high $M_*/M_{gas}$ systems.
This drives the derived value of $Z_{EEP}$ to nonphysical negative values.
For this reason, we recalculate the expected contribution from $Z_*$.
We assume two values are known: (1) no metals will be produced for a system with $M_*/M_{gas}=0$, (2) the highest stellar fraction system (NGC1132 with $M_*/M_{gas}=0.48\pm0.09$) has $Z_* = Z_{ICM}$ (\citetalias{Loew2013}).
From this we derive $Z_*=(0.63 \pm 0.12)(M_*/M_{gas})$.

The value of $Z_{EEP}$ is then found by subtracting $Z_*$ from $Z_{ICM}$, resulting in $Z_{EEP}=(-0.89 \pm 0.12) (M_*/M_{gas}) + (0.46 \pm 0.01)$, valid over the range of $0.035 < M_*/M_{gas} < 0.43$.
This derived trend is the basis for further calculations, as we are able to determine $Z_{EEP}$ for each system.

We intend to further explore the relationship between $Z_{ICM}$, $Z_{*}$, $M_*/M_{gas}$, and other properties of galaxy groups and clusters in a future work.

\section{EEP Assumptions}
\label{sec:assumptions}
In order to move forward with calculations pertaining to EEP, we must make assumptions about key quantities.
We discuss the major assumptions made in the following list:
\begin{enumerate}
    \item We adopt a $\Lambda$CDM cosmology with $H_0 = 70$ km s$^{-1}$ Mpc$^{-1}$, a baryon density parameter of $\Omega_b = 0.046$, a dark matter density parameter of $\Omega_{DM} = 0.24$, and a matter density parameter of $\Omega_m = 0.32$ \citep{Planck2018cosmo}.

    \item The relative Fe mass yields for SN Ia and SNcc adopted are $y_{Ia}=0.743 M_\odot$ and $y_{CC}=0.0825 M_\odot$. These values follow from \citetalias{Loew2013} and \cite{Kobayashi2006}.

    \item We adopt an expected $Z_{ICM}$ stellar contribution of $Z_* = (0.63 \pm 0.12) (M_*/M_{gas})$, as derived in Section \ref{sec:survey}.

    \item We adopt four IMFs (Figure \ref{fig:imfs}), discussed further in Section \ref{sec:imf}. Each IMF chosen has been shown to reproduce $Z_{ICM}$ \citep{Loew2013, Corazza2022}.

    \item We assume an extended star formation at z=10, as discussed in \cite{Freund2021}. These assumptions are supported by findings of cosmological simulations of dwarf galaxies which assume high-z star formation \citep{Brown2014, Jeon2021}, and find that burstiness of star formation history has little impact on simulation results \citep{Weisz2014}.

    \item We adopt a progenitor mass range for SN Ia of 3-8 $M_\odot$ and SNcc of 8-95 $M_\odot$. Both \citetalias{Loew2013} and \cite{Maoz2012} assumed a mass range of 3-8 $M_\odot$ for SN Ia progenitors. However, recent modeling of Type Ia systems suggests that the lower limit could be $\sim$1.8$M_\odot$ \citep[e.g.][]{Anton2020}. With the IMFs adopted, the percentage of stars within the 2-3 $M_\odot$ range from 1.36\%-12.28\%. We proceed with a conservative approach and adopt a lower mass limit of 3 $M_\odot$, potentially underestimating the number of expected SN Ia. For further discussion and recent on SN Ia progenitors, we point the reader to \cite{Lui2023}. The upper limit for SNcc requires consideration of the low-mass cutoff for pair-instability supernovae (PISN), which is 65-140 $M_\odot$ \citep{Heger2002, Chatz2012, Morsony2014}. From \cite{Woosley2017}, a star with an initial mass of 95 $M_\odot$ will result in a PISN producing 0.0045 $M_\odot$ of Fe. However, the yield of PISN, and upper mass bound may be higher \citep{Jeon2015, Karlsson201}. The yields in these works vary as 0.01-0.5 $M_\odot$, and the upper mass cutoff reaches $\sim$260 $M_\odot$. The mass range 95-150 $M_\odot$ comprises $3.5 \%$ of a diet-Salpeter IMF. If we assume $y_{PISN}$ of 0.0045 $M_\odot$, 0.01 $M_\odot$, and 0.5 $M_\odot$, the PISN contribute only 0.07\%, 0.18\% and 8.1\% respectively, to the total metal yield from SN Ia, SNcc and PISN. This is a negligible amount. If we adopt an upper limit of 260 $M_\odot$, we find 6\% of the total mass resides in 95-260 $M_\odot$. This produces fractional contributions to the total metal yield of 0.16\%, 0.35\%, and 15\%, respectfully. We choose a PISN lower mass bound of 95 $M_\odot$ to be within the uncertainty of SNcc versus PISN cutoff, while accounting for a majority of the stellar population and Fe production. For this work, we continue with an IMF upper mass bound of 150 $M_\odot$ and do not consider PISN.

\end{enumerate}

\section{Initial Mass Function}
\label{sec:imf}

\begin{figure}
    \centering
    \includegraphics[width=1\linewidth]{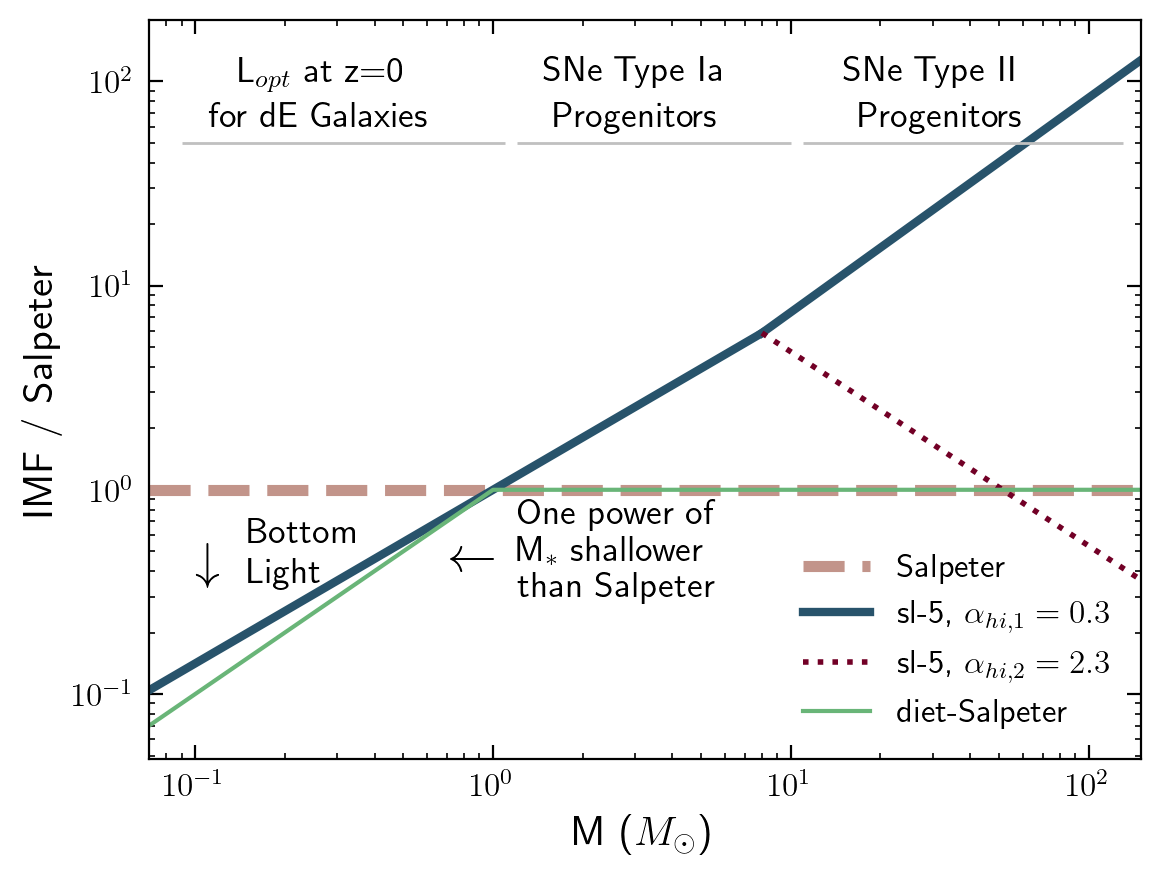}
    \caption{The EEP IMF may be constrained in mass ranges through various measurements. Possible EEP IMFs shown are \textit{sl-5} from \citetalias{Loew2013} (solid blue and dotted purple lines), and diet-Salpeter (as adopted in \cite{Blackwell2022}). Both are divided by Salpeter (dashed light purple line). The IMFs are described in Table \ref{tab:imfs}.}
    \label{fig:imf_analysis}
\end{figure}

The current initial mass function (IMF) of the EEP is unknown, yet a few constraints are available (described in Figure \ref{fig:imf_analysis}).
The IMF must be bottom-light, with a higher low-mass cutoff, in order to produce the amount of metals measured in the ICM without being visible. 
The formation time period must also align with observations of SN Ia, otherwise the total number of SN Ia would exceed observations if the stellar formation is too close to present day.

Possible IMFs for the EEP population have been discussed in \citetalias{Loew2013} and \cite{Corazza2022}.
However, these IMFs were derived with the goal of recreating the measured metal abundance only.
Here, we include other observational constraints, such as the SN Ia rate as a function of redshift and the observed luminosity function of galaxy groups and clusters, to better constraint the EEP IMF and time of formation. 
SN Ia contribute approximately 80\% of the metal yield from supernovae, dominating metal production of the EEP.
Enough high-mass stars must be formed to produce SN Ia, but the total number of SN Ia required to produce $Z_{EEP}$ must not exceed the rate observed in galaxy groups and clusters.
If the EEP formed pre-reionization, we expect dE galaxies to be the primary hosts of the EEP.
The luminosity of the remaining low mass ($<$1$M_\odot$) stars must not exceed the luminosity of dE galaxies observed in LFs. 

\begin{figure}
    \centering
    \includegraphics[width=1\linewidth]{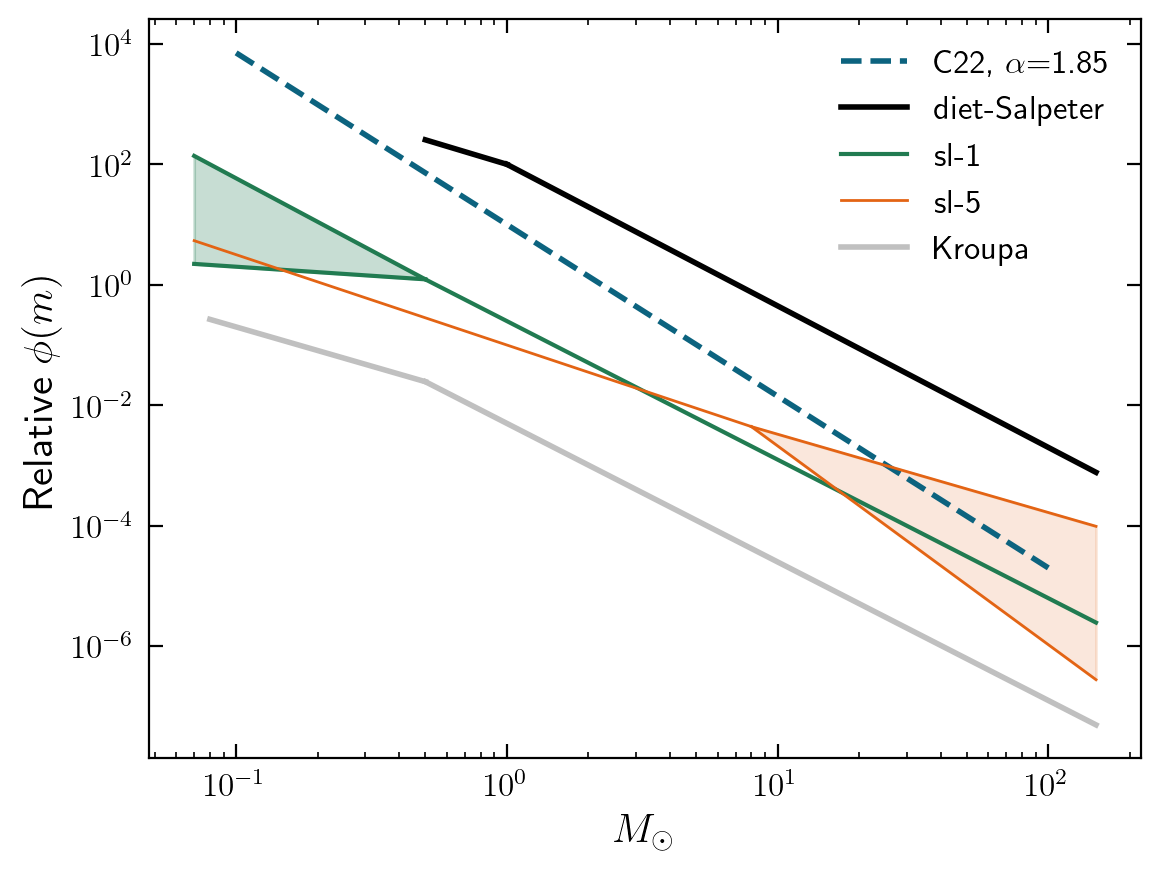}
    \caption{Adopted IMFs for calculations shown compared to a Kroupa IMF (grey line). The normalization of the IMFs have been offset for better visualization. The shaded green and orange regions show the variations in the IMF slope allowed for those mass ranges and derived IMFs from \citetalias{Loew2013}.}
    \label{fig:imfs}
\end{figure}

There are three observational constrains for the IMF of the EEP: (1) the Fe mass of the EEP ($Z_{EEP}$), (2) the SN Ia rate, (3) the luminosity function (LF) of dEs.
We assume various IMFs (Table \ref{tab:imfs}) to perform calculations aimed at determining the expected measurable quantities which can be compared to observational values.
There are a total of 14 IMFs, that recreate $Z_{ICM}$, developed in \citetalias{Loew2013} and \cite{Corazza2022} from which we pick 4, shown in Figure \ref{fig:imfs}.
These IMFs were chosen to represent the most extreme, and average IMFs that reproduce $Z_{ICM}$.
IMFs \textit{sl-1} and \textit{sl-5} have the lowest and highest possible mass breaks and powerlaw-slope values (\citetalias{Loew2013}).
The diet-Salpeter IMF was chosen as a continuation from \cite{Blackwell2022}.
The IMF from \cite{Corazza2022} is the only one determined from simulations and has the highest powerlaw-slope.
All IMFs take the functional form $\Phi(m) = m^{-(1+\alpha)}$.
The four IMFs are shown in Figure \ref{fig:imfs}, compared to the well-known Kroupa IMF.

\begin{table*}[]
    \centering
    \begin{tabular}{c|c|c|c}
         Name & $\alpha$ & Mass Range ($M_\odot$) & Reference  \\
         \hline
         diet-Salpeter & 0.35 & 0.5-1 & \cite{Blackwell2022} \\
                      & 1.35 & 1-150 & \\
         C22 & 1.85 & 0.1-100 & \cite{Corazza2022} \\
         sl-1 & -0.7-1.4 & 0.07-1 & \citetalias{Loew2013} \\
              & 1.3 & 1-150 & \\
         sl-5 & 0.5 & 0.07 - 8 & \citetalias{Loew2013} \\
              & 0.3-2.3 & 8-150 \\
        Kroupa & 0.3 & 0.08-0.5 & \cite{Kroupa2001} \\
               & 1.3 & 0.5-150 & \\
    \end{tabular}
    \caption{Initial mass functions adopted for calculations. Each IMF has been shown to properly recreate $Z_{ICM}$. Kroupa is included for comparison.}
    \label{tab:imfs}
\end{table*}

\section{Type Ia Supernovae}
\label{sec:SNIa}

Supernovae Type Ia can be seen long after the formation of the progenitor stellar population.
The number and distribution of observed SN Ia are heavily influenced by the period and shape of star formation, as well as the number of stars formed within the SN Ia progenitor mass range ($3-8 M_\odot$).
The later is entirely dictated by the IMF of the initial stellar population. 
Therefore, the number and distribution of SN Ia observed can be used to constrain the IMF.
If the IMF slope within the mass range $3-8 M_\odot$ is shallower, more SN Ia will be produced and subsequently observed compared to an IMF with a steeper slope within the same mass range. 
In this section we discuss the current constraints on the IMF using the observed distribution of SN Ia over cosmic time within galaxy clusters.

\subsection{Delay Time Distribution}
The number of SN Ia as a function of redshift since formation can be described by the delay time distribution (DTD) over the redshift range $\sim$10-0 \citep{GalYam2004}.
The DTD, Equation \ref{eq:dtd}, has been characterized in galaxy groups and clusters beginning with a starburst assumption and nearby systems, and it has subsequently evolved to include various star formation scenarios and observations of SN Ia rates in systems out to $z=1.75$ \citep[e.g.][]{Maoz2010, Freund2021}.
\begin{equation}
    DTD(t) = R_1\big( \frac{t}{Gyr} \big)^\alpha
\label{eq:dtd}
\end{equation}
More recently, the constraint on the star formation history has been relaxed to extended star formation, yielding an amplitude of $R_1 = 0.41^{+0.12}_{-0.10} \times 10^{-12}$ yr$^{-1}$ M$_\odot^{-1}$ and a power-law index $\alpha = -1.09^{+0.15}_{-0.12}$ \citep{Freund2021}.

We use the derived DTD shape from \cite{Freund2021} to recreate the expected EEP DTD. 
We do this by determining $Z_{EEP}$ for a single system, calculating the number of SN Ia required to produce $Z_{EEP}$, and normalizing the integrated DTD to the total number of SN Ia required, as done in \cite{Blackwell2022}.
We choose galaxy cluster A2261 for this analysis as its stellar fraction, $0.095\pm 0.029$, and fit $Z_{ICM}$, $0.44 \pm 0.04$ $Z_\odot$, represent the median of the survey.
The value $Z_{EEP}$ is determined by subtracting $Z_*=(0.63 \pm 0.12)(M_*/M_{gas})$, from $Z_{ICM}=0.44 \pm 0.03$ $Z_\odot$, as described in Section \ref{sec:survey}.
We find $Z_{EEP} = 0.38 \pm 0.04$ $Z_\odot$.
The total EEP Fe mass, $M_{Fe,EEP}$, is found by multiplying $Z_{EEP}$ by the baryonic mass of A2261, $M_{bary}=1.06(\pm0.034)\times 10^{14}$ $M_\odot$, and fractional value of Fe in the sun ($Z_\odot^{Fe}=$0.0018) to find $M_{Fe,EEP}=7.7(\pm0.8)\times 10^{10}$ $M_\odot$ \citep{Grevesse1999, Asplund2009, Lodders2021}.

The total number of SNe (Ia + CC) is found by dividing $M_{Fe,EEP}$ by the average SNe Fe yield, Equation \ref{eq:yield}.
The relative SNe yields are assumed values of $y_{Ia}=0.743 M_\odot$ and $y_{CC}=0.0825 M_\odot$ \citep{Kobayashi2006}.
However, the fractional occurrences of each SNe ($f_{ia}$ and $f_{CC}$) are dependent on the assumed IMF, and can be determined by integrating the IMF over a predefined progenitor mass range for each pathway.
We note that using the ratio of $\alpha$/Fe elements within the ICM may also allow for further constraints of the available IMFs, however, that is outside the scope of this paper.

\begin{equation}
    y_{SNe} = (y_{Ia} \times f_{ia}) + (y_{CC} \times f_{CC})
\label{eq:yield}
\end{equation}

\begin{figure}
    \centering
    \includegraphics[width=1.0\linewidth]{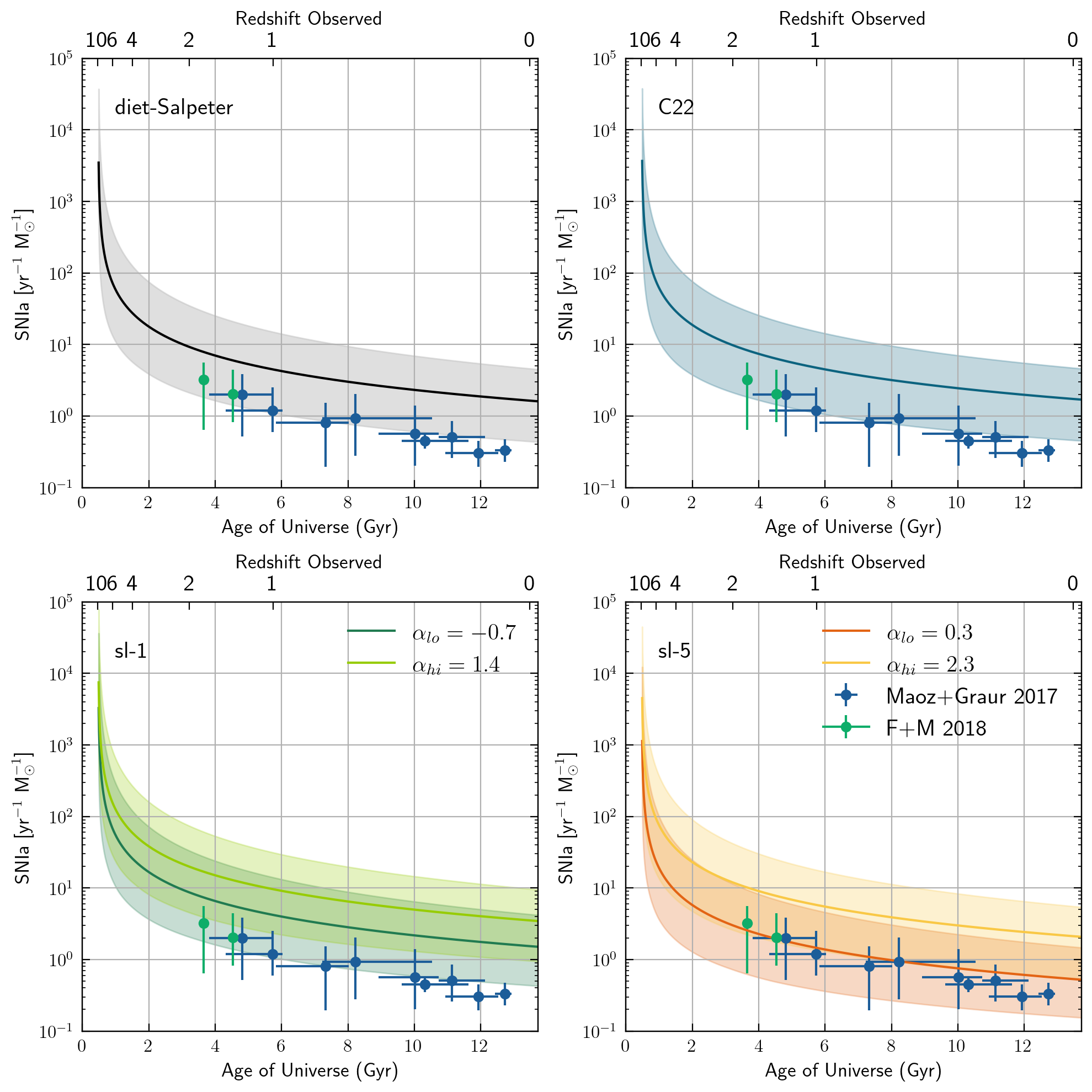}
    \caption{The DTD derived for each IMF. The trends are compared with data from \cite{Friedmann2018} (F+M 2018) and \cite{Maoz2017}. Three of the DTDs overpredict the rate of SN Ia, however, they are still within the 90th percentile (the shaded regions). The best match is for the $\alpha_{hi,1}=0.3$ version of \textit{sl-5} (dark orange line of bottom-right plot).}
    \label{fig:dtd_data}
\end{figure}

We calculate the required number of SN Ia and SNcc by multiplying the total number of SNe by $f_{ia}$ and $f_{cc}$.
We then use the total number of SN Ia to normalize the integrated DTD powerlaw ($R_1$) from \cite{Friedmann2018}.
The derived DTD for each of the IMF, shown in Figure \ref{fig:dtd_data}, is also compared to SN Ia rates in clusters from \cite{Freund2021} and \cite{Maoz2010}.
The DTD errors are determined by sampling $\alpha$ and $R_1$ values weighted by the respective probability density functions. 

The predicted DTDs in Figure \ref{fig:dtd_data} fall within observational errors, but are consistently high. 
Variations within the \textit{sl-5} high-mass slope can match observations (bottom-right plot of Figure \ref{fig:dtd_data}). 
To better understand why three of the DTDs consistently overpredict the number of observed SN Ia, we consider the fractional amount of stars within the SN Ia progenitor mass range that produce SN Ia, $F$.
For the diet-Salpeter IMF specifically, if $F\sim0.5$ (50\% of the stars within $3-8$ $M_\odot$ produce SN Ia), then we find agreement between the predicted DTD and observed SN Ia rates.
In our initial calculation, we assume all stars within the mass range 3-8 $M_\odot$ produce SN Ia. 
This is unlikely, and there are several uncertainties regarding which stars follow the pathway, and which specific pathway, of SN Ia. 
Due to this, there is a poorly known fractional amount of stars within a progenitor mass range that become SN Ia. 
For example, if we assume a lower mass limit for SN Ia of $1 M_\odot$, the fractional stellar mass increases from 19\% in $3-8 M_\odot$ to 51\% in $1-8 M_\odot$.
$F$ should increase by approximately the same amount, 32\%.
There are other assumptions that go into this calculation, as noted in Section \ref{sec:assumptions}, that undoubtedly play into the overprediction of the total number of SN Ia required.

\subsection{Best Fit IMF}
\label{sec:imf_mcmc}

\begin{table*}[]
    \begin{centering}
    \begin{tabular}{c | c | c | c | c | c | c} 
        IMF & $M_{br}$ ($M_\odot$) & $\alpha_{lo}$ & $\alpha_{hi}$ & $F$ & $f^{Ia}$ & $f^{CC}$ \\
        \hline
         Low $M_{br}$ & 1.75 & $-0.13 \pm 0.24$ & $0.53 \pm 0.01$ & & $0.14\pm0.01$ & $0.86 \pm 0.01$ \\
         & 1.75 & $0.20^{+0.74}_{-0.61}$ & $0.96^{+0.59}_{-0.33}$ & $0.47^{+0.22}_{-0.27}$ & $0.27 \pm 0.10$ & $0.73 \pm 0.10$ \\
         High $M_{br}$ & 6 & $1.06 \pm 0.11$ &  $0.53 \pm 0.01$  &  & $0.10 \pm 0.01$ & $0.90 \pm 0.01$\\
         & 6 & $0.18^{+0.76}_{-0.61}$ & $1.15^{+0.74}_{-0.48}$ & $0.44^{+0.17}_{-0.27}$ & $0.56 \pm 0.23$ & $0.44 \pm 0.23$ \\
    \end{tabular}
    \caption{Derived best fit IMF parameters and fractional values of SN Ia and SNcc. The slope values presented are for an IMF with functional form $\Phi(m) = m^{-(1+\alpha)}$ (per unit log mass). Where a value of $F$ (fraction of the progenitor mass range $3-8 M_\odot$ that become SN Ia) is not defined, it was assumed $F=$1 and not a free parameter in the MCMC.}
    \label{tab:best_imfs}
    \end{centering}
\end{table*}

Here, we investigate the best-fit IMF parameters that derive a DTD most closely aligning with the available data.
There are three primary variables in the IMF: the low-mass slope ($\alpha_{lo}$), the high-mass slope ($\alpha_{hi}$), and the mass break between the two slopes ($M_{br}$).
Ranges of each parameter are determined in \citetalias{Loew2013} as these IMFs are shown to recreate $Z_{ICM}$: $-0.7 < \alpha_{lo} < 1.4$, $0.5 < \alpha_{hi} < 2.3$, and $0.5 < M_{br} < 8$.
We ran a Markov Chain Monte Carlo (MCMC) to determine the best-fit IMF parameters, using the listed variable ranges as hard prior limits to avoid any deviations that could result in non-physical $Z_{ICM}$ (\citetalias{Loew2013}).
Fitting was done in log-space, and the asymmetric errors on observational data from \cite{Maoz2010} and \cite{Friedmann2018} were averaged.

\begin{figure}
    \centering
    \includegraphics[width=1.0\linewidth]{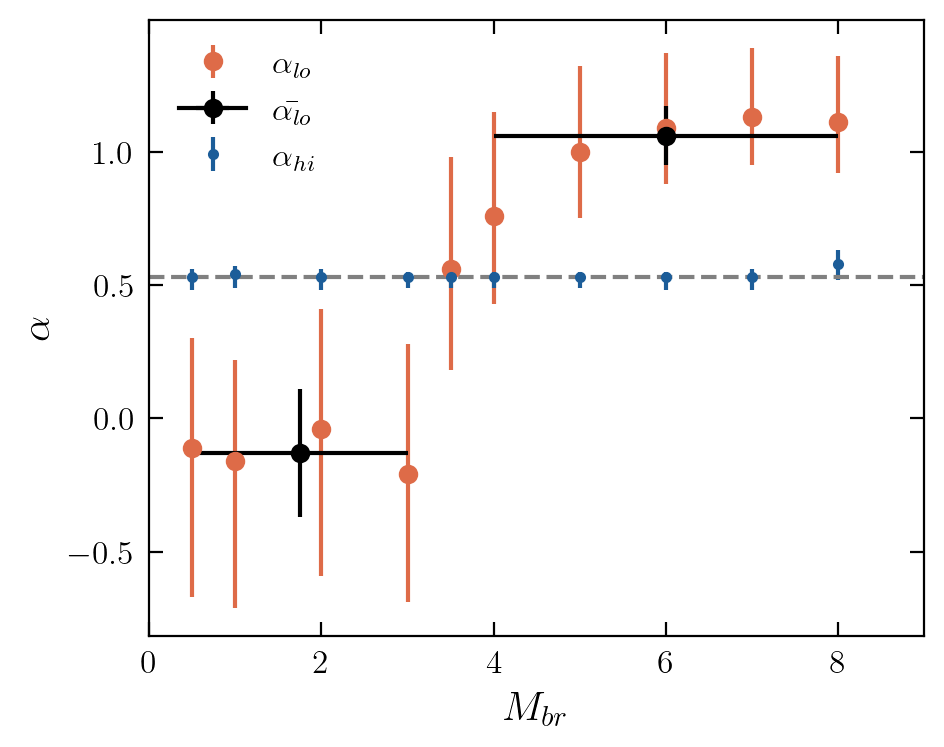}
    \caption{Fit values of $\alpha_{lo}$ (red data points) and $\alpha_{hi}$ (blue data points) with fixed values if $M_{br}$. The value for $\alpha_{hi}$ is consistent within 1$\sigma$ at $0.53 \pm 0.01$ (dashed grey line) over the entire $M_{br}$ range. We find a weighted average value of $\alpha_{lo}=-0.13 \pm 0.24$ within $0.5 < M_{br}/M_\odot < 3$, and $\alpha_{lo}=1.06 \pm 0.11$ within $4 < M_{br}/M_\odot < 8$.}
    \label{fig:alpha_mbr}
\end{figure}

\begin{figure}
    \centering
    \includegraphics[width=1\linewidth]{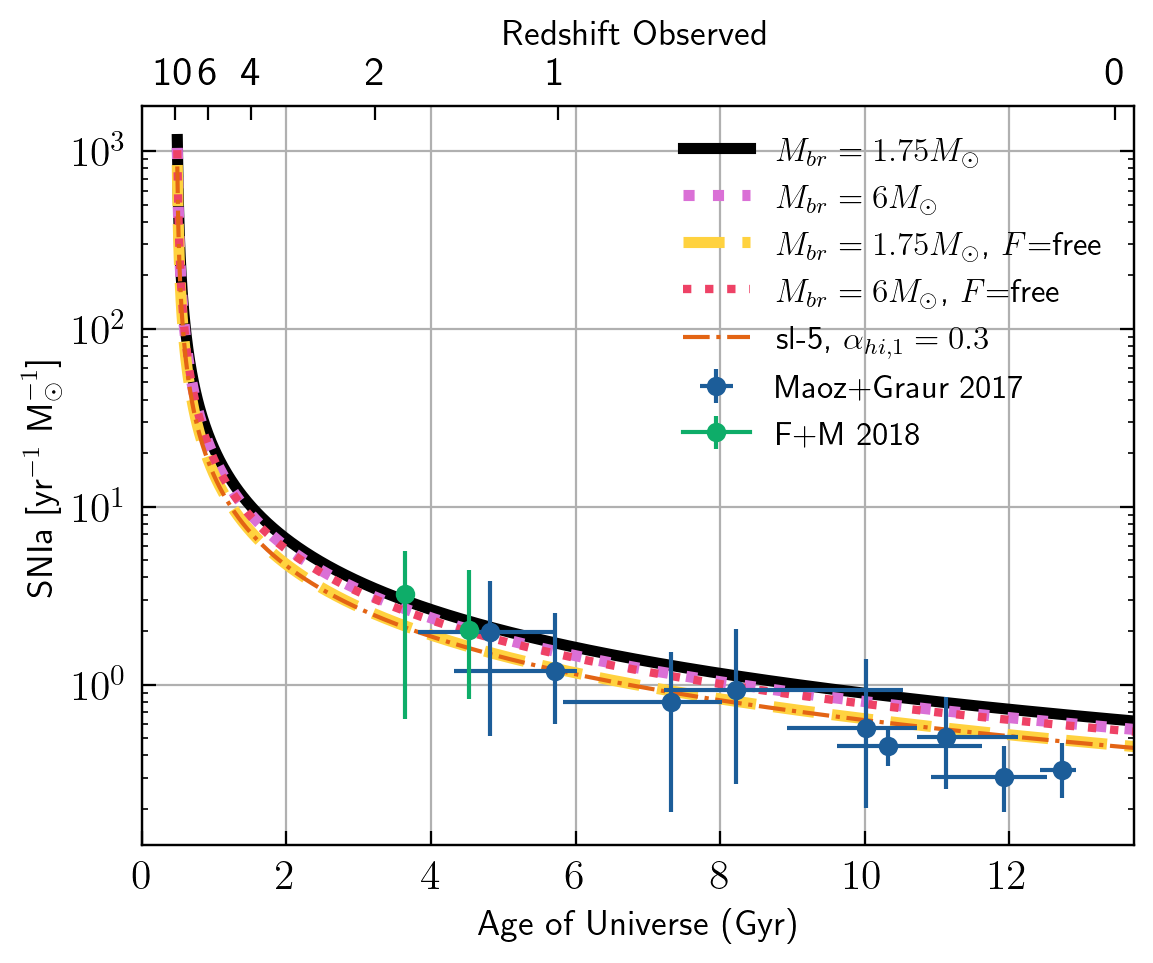}
    \caption{Best fit IMFs determined with an MCMC with $M_{br}=1.75 M_\odot$ (black line), and $M_{br}=6 M_\odot$ (purple dotted line). We allowed $F$ to be fit as a free parameter in the MCMC and show the fit results with $M_{br}=1.75 M_\odot$ fitting $F=0.47^{+0.22}_{-0.27}$ (yellow dashed line), and $M_{br}=6 M_\odot$ fitting $F=0.44^{+0.17}_{-0.15}$ (red dotted line). For comparison, we plot \textit{sl-5} with $\alpha_{hi,1}=0.3$ from \citetalias{Loew2013} (orange dash-dot line) which agrees with the $M_{br}=1.75 M_\odot$ and $F=0.47^{+0.22}_{-0.27}$ fit. All fit results plotted are summarized in Table \ref{tab:best_imfs}.}
    \label{fig:mcmc_imf}
\end{figure}

During fitting, we identified a degeneracy between $M_{br}$ and $\alpha_{lo}$.
We held $M_{br}$ at constant values within the prior mass range and fit for $\alpha_{lo}$ and $\alpha_{hi}$, Figure \ref{fig:alpha_mbr}.
The value for $\alpha_{hi}$ is consistently fit within $1\sigma$, with a weighted average of $0.53 \pm 0.01$.
There are two $M_{br}$ ranges within which $\alpha_{lo}$ is fit within $1\sigma$.
In $0.5 < M_{br}/M_\odot < 3$, we find a weighted average of $\alpha_{lo}=-0.13 \pm 0.24$, and in $4 < M_{br}/M_\odot < 8$, the weighted average increases to $\alpha_{lo}=1.06 \pm 0.11$.
We summarize the mass ranges as the median values for future use and simplicity, 1.75$M_\odot$ and 6$M_\odot$.
Notably, the fit values associated with $M_{br}=6 M_\odot$ produce an IMF that is closer to the one seen at present day \citep{Salpeter1955}.
There is a transitional value of $\alpha_{lo}$ within $3 < M_{br}/M_\odot < 4$, which we exclude from any averages.
We proceed with calculations and DTD derivations using the best-fit IMF in each $M_{br}$ regime, summarized in Table \ref{tab:best_imfs}, and show the derived DTDs in Figure \ref{fig:mcmc_imf}.

Our best constrained IMF parameters can be compared to a final observable value - the fractional amount of SN Ia to SNcc ($f^{Ia}$ and $f^{CC}$, respectively, Table \ref{tab:best_imfs}) \citep[see also][]{Bulbul2012}.
Our determined $f_{Ia}$ is lower than what is found in the literature. 
\citetalias{Loew2013} finds $f_{Ia}=20\%$ using a diet-Salpeter IMF. 
\cite{Bulbul2012} use the spectral fitting model \textit{snapec} to determine $f_{Ia}$ in the range 30($\pm5.4$) to 37($\pm7.1$)\%, however, this model assumed a Salpeter IMF.
Studies of the ICM metallicity find values similar to ours, 29-45\% \citep{Mernier2016}, while observations derive 8-15\% \citep{Mannucci2005} in conflict with our findings.
Notably, the variations in IMF \textit{sl-5} from \citetalias{Loew2013} derive a $f^{Ia} = 7-97\%$ spanning the range of $f^{Ia}$ in the literature, and the best-fit values for $\alpha_{lo}$ and $\alpha_{hi}$.
The model that most closely fits the literature is $M_{br}=1.75 M_\odot$, $\alpha_{lo}=0.20^{+0.74}_{-0.61}$, $\alpha_{hi}=0.96^{+0.59}_{-0.33}$, and $F=0.47^{+0.22}_{-0.27}$, resulting in $f^{Ia} = 0.27 \pm 0.10$.

\subsection{Fraction of stars that result in SN Ia}
\label{sec:fudge_factor}

Here we consider adding $F$, the fractional amount of stars within the progenitor mass range $3-8 M_\odot$ that produce SN Ia, in our fittings.
We originally assumed this value to be 1, a likely incorrect assumption.
We introduce $F$ as a free parameter in our MCMC fit, and calculate the best-fit values for $F$, $\alpha_{lo}$, and $\alpha_{hi}$, holding $M_{br}=1.75 M_\odot$ and $M_{br}=6 M_\odot$. 

The best fits, summarized in Table \ref{tab:best_imfs}, recreate the number of SN Ia required to produce $Z_{ICM}$. 
The derived DTDs, shown in Figure \ref{fig:mcmc_imf}, are also in agreement with those obtained without allowing $F$ to be a free parameter.
However, when $F$ is included as a free parameter, the error bars on the fit variables are approximately twice as large as those from any single MCMC fit that assumes $F=1$.
This indicates that we have insufficient data to confidently constrain the three IMF parameters and, therefore, do not consider these IMFs as the best fit.

We can compare the fit value of $F$ with a theoretical and more physically motivated calculation. 
Here, we present an estimated calculation, but recommend conducting a comprehensive stellar evolutionary analysis to determine the true value of $F$. 

Three primary values are needed for this estimation: the fraction of stars in binary systems, the fraction of binary systems with a period short enough to allow for mass transfer, and the fraction of binary systems within the mass range that can form a binary without destroying the companion and still permit mass transfer.
Multiplying each of these three factors together gives the fractional amount of stars within the progenitor mass range that result in SN Ia.

Approximately 54\%$\pm$3\% of solar-type stars are single \citep[assumed from a study of the Milky Way,][]{Raghavan2010}.
We assume the remaining 46\% of the stars to be in binary systems.
The orbital period distribution can be described as Equation \ref{eq:orbital}, where $P$ is the orbital period in days, $\overline{log P}$=4.8, $\sigma_{logP}$=2.3, and $Cst$ is a normalization constant \citep{Duquennoy1991}.
Figure 1 in \cite{Moe2017} shows the expected period for binary systems that may turn into SN Ia, $2 < logP < 3.7$.
We integrate Equation \ref{eq:orbital} over the total binary population range, $0 < log P < 10$, and that expected for SN Ia, finding the fractional amount of binary systems with $2 < logP < 3.7$ to be 0.21 (21\%).

\begin{equation}
f(log P) = Cst \: \text{exp}\Big[\frac{-(log P - \overline{log P})^2}{2\sigma^2_{logP}}\Big]
\label{eq:orbital}    
\end{equation}

We use Equation \ref{eq:mass_dist} to determine the fractional number of binaries within a defined range of mass fractions, $q = M_2/M_1$, where $\mu = 0.23$, and $\sigma_q = 0.42$ \citep{Duquennoy1991}. 
From Figure 1 in \cite{Moe2017}, we find $q$ values that may result in SN Ia as $0.1 < q < 0.3$.
With this assumption, we determine the fractional number of binaries with $0.1 < q < 0.3$ as 0.28 (28\%).

\begin{equation}
    f(q) = k \: \text{exp}\Big[\frac{-(q - \mu)^2}{2\sigma^2_q} \Big]
\label{eq:mass_dist}    
\end{equation}

The resulting fractional amount of stars within the progenitor mass range to result in SN Ia is: $0.44 \times 0.21 \times 0.28 = 0.03$ (3\%).

This is an order of magnitude smaller than our MCMC determined $F$ of $\sim 0.45$, which is needed to match our predictions with observations of SN Ia.
\cite{Maoz2008} and \cite{Mannucci2008} both provide similar $F$ estimates of 2-40\% using various IMF assumptions, with \cite{Maoz2008} concluding 15\% to be most consistent.
The difference in our higher (fit) and lower (calculated) $F$ values may indicate that for the EEP, the IMF and binary separation distribution was different than at low redshifts.

We conclude that our findings of $F$ are poorly constrained. 
Adding $F$ as a free parameter derives a DTD that matches observations, but does not improve the errors on our derived IMF values.
A great deal is still unknown about SN Ia pathways within nearby star systems, let alone at high-redshift as the EEP must have been.

\section{Luminosity Functions}
\label{sec:lf}

The hosts of the EEP are currently unknown, but are likely to be an early stellar population dominated by lower mass galaxies.
Low optical luminosity dwarfs around the Milky Way present with an IMF (really a present-day mass function) that is flatter ($\alpha \sim 0.2$) at masses  $<$1 $M_\odot$ \citep{Geha2015, Gennaro2018Zdwarfs, Hansen2020}.
This is similar to a diet-Salpeter IMF like suggested for the EEP.
A similar, shallow IMF is found in globular clusters \citep{Cadelano2020gcIMF}, and low-mass, low-metallicity systems \citep{202PrgometEDGE}.
In large systems such as galaxy clusters, these low-mass, low-luminosity, and low-metallicity systems are dwarf Elliptical (dE) galaxies.
We see a large population of these dE galaxies in galaxy clusters \citep[e.g.,][]{Blanton2005, Popesso2005, Trentham2005, Yamanoi2012}.
Here we posit these systems to be the original hosts of the EEP, and investigate the consequences of different IMFs on the current state (remaining light) of the dE galaxies.

The luminosity function (LF) of galaxy clusters describes the number of galaxies at a given magnitude.
It is best described by a double Schechter function, each with a form described by Equation \ref{eq:schechter}.
Here, $L_*$ is the turnover luminosity for a component, $\beta$ is the slope, and $n_*$ is the space density normalization.

\begin{equation}
    \Phi(L)dL = n_*(L/L_*)^\beta exp(-L/L_*)d(L/L_*)
\label{eq:schechter}
\end{equation}

The LF for field galaxies, the Local Group, and nearby groups can be described by a single Schechter component with a slope of $\beta\approx 1.2$ over M(R)=-20 to M(R)=-10 \citep{Lin1996, Phillipps1998, Trentham2002, Blanton2003}, whereas, the LF for galaxy clusters is unique.
The low-luminosity component in galaxy groups and clusters is dominated by dE galaxies.
If one posits that these dE galaxies are some of the original hosts of the EEP, additional constraints may be placed on the IMF of the EEP.
That is, for an assumed IMF, does the remaining light of the low-mass EEP stars exceed observations of dE galaxies in LFs?
We investigate the importance of the dE galaxies and remnants of the EEP population by studying the LF of the Coma cluster.

\subsection{Observed Luminosity}
The Coma cluster luminosity function has been studied extensively over the past several decades in multiple filters and fields \citep[i.e.,][]{Bernstein1995, Trentham1998, Adami2000, Andreon2002, Mobasher2003, Milne2007, Yamanoi2012}.
Despite the many studies, there is still disagreement about the slope of the luminosity function for Coma, as well as indication for a steeper low-luminosity slope \citep{DePropris1995}.
Differences between the studies may account for the inconsistencies.
Some studies use spectroscopic redshift to identify cluster members, limiting the number of identifiable low-luminosity systems \citep{Adami2000, Mobasher2003}.
\cite{Mobasher2003} covers $-23<M_R<-16$ and derives $\beta=-1.18^{+0.04}_{-0.02}$ while \cite{Milne2007} finds $\beta = -2.29\pm 0.33$ over $-14.58<M_R<-9.08$.
The differences among these two magnitude ranges indicate two different powerlaw slopes for the high- and low-luminosity regimes.

\begin{figure}
    \centering
    \includegraphics[width=1\linewidth]{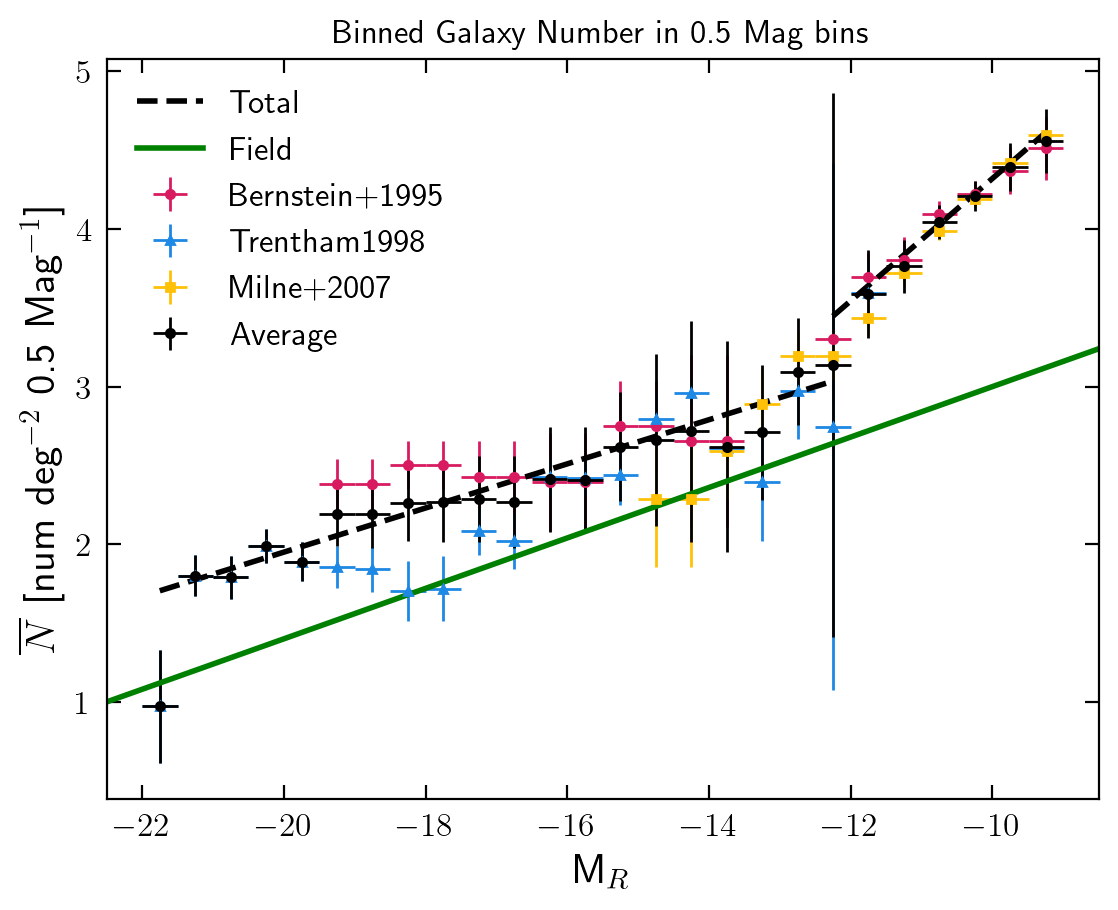}
    \caption{Luminosity function of Coma cluster from \cite{Bernstein1995}, \cite{Trentham1998}, and \cite{Milne2007} (red, blue, and yellow data points respectively). The averaged Coma LF is shown as the black points. We derive a powerlaw slope of $\beta=-1.35\pm0.06$ for $-22<M(R)<-12$ and $\beta=-1.97\pm0.23$ for $-12<M_R<-9$ (the black dashed lines). There is a break in slope at $M_R\sim-12$. We plot the fit LF function trend for field galaxies with $\beta=-1.4$ (solid green line) for comparison \citep{Blanton2005}. There is an excess of low-luminosity galaxies in the Coma LF compared to the field. All counts have been normalized to a 1$^\circ$ field.}
    \label{fig:lum_func}
\end{figure}

We identify three works to analyze measurements, average, and fit a total LF of the Coma cluster: \cite{Bernstein1995}, \cite{Trentham1998} and \cite{Milne2007}.
These papers cover a large range of magnitude in the R band ($-22<M_R<-9$), provide the apparent magnitude, and publish galaxy counts in comparable magnitude bins.
The absolute magnitude is calculated by adopting a distance modulus of 34.83 \citep{Trentham1998, Milne2007}.
Figure \ref{fig:lum_func} shows the three LFs normalized to a 1$^\circ$ field.
We group the published counts into 0.5 magnitude bins, propagating errors, to determine an average LF for Coma, Figure \ref{fig:lum_func}.

We follow the fitting method outlined in \cite{Yamanoi2012} to find the slope of the LF for the high- and low-luminosity components for Coma. 
The logarithmic slope of the luminosity function can be fit using linear regression and Equation \ref{eq:beta} \citep{Andreon2002, Yamanoi2012}.
\begin{equation}
    log\phi = -0.4(\beta + 1)M + C
\label{eq:beta}
\end{equation}
We use an MCMC to fit the logged luminosity function for $\beta$, and the constant, C.
Bootstrapping was used to determine the magnitude break between the high- and low-luminosity components as $M_R\sim-12$.
The fit values were found to be $\beta=-1.35\pm0.06$ and $C=4.75\pm0.46$ for $-22<M_R<-12$ and $\beta=-1.97\pm 0.23$ and $C=8.20^{+0.98}_{-0.97}$ for $-12<M_R<-9$.

Absolute magnitude was converted to luminosity using a solar R-band absolute magnitude of 4.42.
We then re-scale the 1$^\circ$ normalized field as if it were observed within $R_{500}=47\arcmin$ of Coma \citep{Planck2013}.
We calculated all values of $Z_{EEP}$ within $R_{500}$.
For consistency, we choose to rescale the LF study to $R_{500}$.
A Reimann sum was used to determine the cumulative luminosity for Coma in R-band within $R_{500}$ finding $9.14(\pm1.94)\times 10^{12}$ $L_\odot$.
For the high-luminosity galaxies ($M_R<-12$) the cumulative luminosity was calculated to be $L_{hi}=8.98(\pm1.93)\times 10^{12}$ $L_\odot$, and $L_{lo}=1.63(\pm0.62)\times 10^{11}$ $L_\odot$ for low-luminosity galaxies ($M_R>-12$).

\subsection{Expected EEP Luminosity}

The primary question is whether the remaining EEP luminosity exceeds the observed dE luminosity in galaxy clusters.
With the four IMFs described in Table \ref{tab:imfs} and the best-fit IMFs in Table \ref{tab:best_imfs}, we calculate the EEP cumulative luminosity in the remaining stellar population, and compare that to the cumulative luminosity in $M_R > -12$ from the Coma LF.

The following steps outline how we determine the cumulative luminosity for each:
\begin{enumerate}
    \item For Coma, calculate the missing metal mass that not explained by the visible stellar populations ($M_{Z,EEP}$).

    \item Determine the number of SN Ia required to produce $M_{Z,EEP}$, $N_{Ia}$.

    \item Use $N_{Ia}$ to normalize the IMF over the mass range 3-8 $M_\odot$. Propagate that normalization to the entire mass range. That is, the integral of the IMF over 3-8 $M_\odot$ must be equal to $N_{Ia}$, and the full IMF must be continuous.

    \item Assume the EEP formed at z=10. Stars with mass $<0.9 M_\odot$ still exist today. Convert mass to luminosity using $L \propto M^4$ for stars with mass $<0.9 M_\odot$.

    \item Sum the luminosities of stars with $M<0.9 M_\odot$ to determine the cumulative luminosity at the redshift of Coma, $L_{EEP}$ for a given IMF.
\end{enumerate}

\begin{table}[]
    \centering
    \begin{tabular}{c|c|c}
    IMF & $\mathcal{R}_{lo}$ & $\mathcal{R}_{hi}$ \\
    \hline
        dS & 199.1\% & 1.6\% \\
        C22 & 973.5\% & 14.1\% \\
        sl-1$_{lo}$ & 271.9\% & 2.8\% \\
        sl-1$_{hi}$ & 692.1\% & 9.6\% \\
        sl-5$_{lo}$ & 15.2\% & $<$0.1\% \\
        sl-5$_{hi}$ & 56.1\% & $<$0.1\% \\
        $M_{br}=1.75 M_\odot$ & 26.2\% & $<$0.1\% \\
        $M_{br}=6 M_\odot$ & 62.2\% & $<$0.1\% \\
    \end{tabular}
    \caption{Percent of EEP luminosity remaining for each IMF, defined in Column 1. Column 2 is the percentage of dE luminosity produced by the remaining EEP population, Equation \ref{eq:llo}. Column 3 is the percentage of remaining EEP light in the high-luminosity ($-22<M(R)<-12$) galaxies if the excess light from Column 2 (anything $>$100\%) merged into the high-luminosity galaxies, Equation \ref{eq:lhi}. $F=1$ in the two IMFs defined by $M_{br}$. The top 4 lines (3 models) are already ruled as unlikely by DTD predictions.}
    \label{tab:imf_calcs}
\end{table}

\subsection{EEP Contribution to Coma LF}

Here we calculate the luminosity contribution from the EEP to the Coma LF. There are several important factors to consider: the fraction of the EEP luminosity to the luminosity of Coma, the fraction of the EEP luminosity to the Coma low-luminosity component ($-12 < M_R < -9$), and the fraction of the EEP luminosity to the Coma high-luminosity component ($-22 < M_R < -9$). 
None of the IMFs exceed the luminosity of Coma, so are not immediately ruled out.

The contribution of the EEP luminosity to the low- and high-luminosity regimes takes additional consideration. 
We posit that the dE galaxies, low-luminosity end of the Coma LF, were the original hosts of the EEP. 
This contribution is calculated as the ratio of the remaining EEP luminosity ($L_{EEP}$) for a given IMF, to $L_{lo}$, Equation \ref{eq:llo}.
\begin{equation}
	\mathcal{R}_{lo} = \frac{L_{EEP}}{L_{lo}}
\label{eq:llo}
\end{equation}
If $\mathcal{R}_{lo}< 100\%$, then all of the remaining EEP luminosity can be accounted for the in the dE population of Coma.

However, some unknown fractional amount of these galaxies and EEP have merged with the high-luminosity galaxies. 
There is an important second measure, $\mathcal{R}_{bri}$ which compares $L_{EEP}$ to the high-luminosity, and largest, part of the LF. 
This fraction, given by Equation \ref{eq:lhi}, is a lower limit as it depends on the IMF of the high-luminosity galaxies and the merging fraction of dE galaxies.
\begin{equation}
	\mathcal{R}_{hi} = \frac{L_{EEP} - L_{lo}}{L_{hi}}
\label{eq:lhi}
\end{equation}
If the IMFs of the EEP and the high-luminosity galaxies are the same, then $\mathcal{R}_{bri}$ is the true fraction of the EEP population in the luminous galaxies. 
However, it is likely the EEP IMF produces fewer low-mass stars than the currently visible dE and high-luminosity galaxies. 
Additionally, a value of $\mathcal{R}_{bri}=0\%$, as is true for some of the IMFs, suggests that none of the dE galaxies have merged with the high-luminosity galaxies.
This is nonphysical as galaxy clusters are incredibly dynamic systems.

Table \ref{tab:imf_calcs} shows the results of these comparisons. 
We cannot rule out a single EEP IMF, nor narrow down the three most likely IMFs (\textit{sl-5}, and the two derived from MCMC fitting). 
C22 produces the most extreme values with a required lower-limit merging fraction of $\mathcal{R}_{bri}=14.1\%$. 
Most of the remaining IMFs will contribute only 3\% or less to the high-luminosity component. 
Theoretical predictions of the fractional dE galaxies merged with high-luminosity cluster members would be beneficial in refining, and potentially eliminating, EEP IMFs.

\section{Summary}

In this work, we present example calculations of observable quantities for the theoretical Early Enrichment Population.
We test four IMFs that can reproduce $Z_{ICM}$ \citep{Loew2013, Corazza2022} and calculate SN Ia DTDs, and the remaining EEP stellar luminosity.
Our findings can be summarized in the following points.

\begin{enumerate}[label=(\roman*)]
    \item We calculate predicted DTDs for various IMFs: diet-Salpeter (dS), C22, \textit{sl-1}, and \textit{sl-5}. The DTDs from dS, C22, and \textit{sl-1} overpredict, but are within $1\sigma$, of the observed SN Ia rate \citep{Maoz2017, Friedmann2018}. IMF \textit{sl-5} with $\alpha_{hi,1}=0.3$ most closely reproduces the observations.

    \item We use an MCMC to find the best fit IMF parameters, $\alpha_{lo}$, $\alpha_{hi}$ and $M_{br}$. We found two mass ranges within which $\alpha_{lo}$ was fit within $1\sigma$. The best fits are summarized in Table \ref{tab:best_imfs}, and derived DTDs in Figure \ref{fig:mcmc_imf}. These best-fit IMFs produce a fractional SN Ia of $14\%\pm0.5\%$ for $M_{br}=1.75  M_\odot$, and $10\%\pm1\%$ for $M_{br}=6 M_\odot$. This is lower than theoretical values such as 20\% from \citetalias{Loew2013} and 29-45\% from \cite{Mernier2016}.

    \item We consider an additional factor, $F$, to account for the fraction of stars within the progenitor mass range 3-8$M_\odot$ that produce SN Ia. We calculate a theoretical value of 0.03, and find an MCMC best fit value of $0.47^{+0.22}_{-0.27}$ and $0.44^{+0.17}_{-0.15}$ with an $M_{br}$ of 1.75$M_\odot$ and 6$M_\odot$, respectfully. The lower boundaries of our calculated $F$ values are in agreement with those from the literature, 2-40\% \citep{Mannucci2008, Maoz2008}.

    \item We constrain the low-mass end of the EEP IMF using the Coma cluster LF. We find that only \textit{sl-5}, $M_{br}=1.75 M_\odot$ ($F=1)$, and $M_{br}=6 M_\odot$ ($F=1)$ produce few enough stars with $M<0.9 M_\odot$ to not exceed the Coma cumulative low-luminosity produced by dE galaxies, $\mathcal{R}_{lo}=13-50\%$. However, the IMFs that overpredict dE luminosity may not be ruled out as we must consider merging of dE systems with the high-luminosity galaxies ($-22<M(R)<-12$). C22 predicts the greatest dE excess, $\mathcal{R}_{lo}=$973\%. If we require $\mathcal{R}_{lo}=100\%$, the remaining $873\%$ can be merged into the high-luminosity cluster galaxies and require $\mathcal{R}_{hi}=14\%$. It is likely that a most of the dE galaxies merged into the larger galaxies, however, the exact amount is currently unknown.
\end{enumerate}

The IMF \textit{sl-5} from \citetalias{Loew2013} with $\alpha_{hi,1}=0.3$, and the two derived best-fit IMFs, seem to be the favorable IMFs as they do not exceed observations.
There are a number of assumptions that may be better constrained with progress in fields such as merger tree histories for clusters, and stellar evolution.
Further observations of dE galaxies and SN Ia in clusters will also help to better constrain current model fits.

\section*{Acknowledgements}
We would like to deeply thank Cameron Pratt, Zhijie Qu, and Rui Huang for their thoughtful discussion, feedback, support over the course of this research project. 
The authors would like to gratefully acknowledge support for this program through the NASA ADAP award 80NSSC22K0481.


\bibliography{main}{}
\bibliographystyle{aasjournal}



\end{document}